\documentclass{article}
\usepackage{graphicx}
\begin{document}
\title{Einstein's Working Sheets and His Search For a Unified Field Theory}
\author{Tilman Sauer\\[0.5cm]
{\small Johannes Gutenberg University Mainz}\\ 
{\small 55099 Mainz, Germany}\\
{\small tsauer@uni-mainz.de}\\[0.5cm]
}
\date{Version of \today}
\maketitle

\begin{abstract}%
The Einstein Archives contain a considerable collection of calculations in the form of working sheets and scratch paper, documenting Einstein's scientific preoccupations during the last three decades of his life until his death in 1955. This paper provides a brief description of these documents and some indications of what can be expected from a more thorough investigation of these notes.
\end{abstract} 
\noindent

\section{Introduction}
\label{intro}

The completion of the general theory of relativity in late 1915 is considered Einstein's greatest and most lasting achievement. Nevertheless, soon after the publication of his gravitational field equations, it became clear that Einstein did not consider it a final, or complete, theory. Instead, he continued to contemplate modifications of the field equations and of the conceptual foundations of the theory. One of these modifications, proposed in 1917, consisted in adding a term containing the cosmological constant that was motivated both by difficulties in the interpretation of the early applications of general relativity to the problem of cosmology as well as by the hope to account for the existence and structure of elementary particles. Indeed, the publication of Einstein's field equations was followed by many efforts, both by Einstein and by his contemporaries, to explore, develop, and understand the conceptual and physical implications of the new theory.

While these efforts of developing what would now be called classical general relativity produced many new insights, the early years of general relativity were also characterized by a prominent and widely shared research program aimed at developing a more generalized theory that would unify the known fundamental forces of gravitation and electromagnetism. Another aspect that was motivating theoretical speculation was a strong perception that the new field theoretical conceptualization of gravitation would constitute or exacerbate a dualism of fields and matter which resulted in a program to overcome this dualism by conceiving of particles as special field configurations. Almost immediately following the conception of the theory and even more so after the advent of quantum mechanics as a new foundational theory, efforts to connect quantum features with the field theoretical foundations of general relativity emerged as a third motivation for further speculation.

These three aspects constituted a major research concern for theoretical physicists and mathematicians, loosely captured by a slogan of finding a unified field theory or, sometimes, a theory of everything. During his lifetime, Einstein was a major proponent of this program and, indeed, expended much effort along these lines of research
\cite{TonnelatM1966Theory},
\cite{PaisA1982Subtle},
\cite{BergiaS1993Attempts},
\cite{VizginV1994Theories},
\cite{GoennerH2004History,GoennerH2014History}
\cite{DongenJ2010Unification},
\cite{SauerT2014Program}.

Einstein's published oeuvre is well-known, as is his correspondence, although the latter is only being published by and by through the efforts of the editorial project of the \emph{Collected Papers of Albert Einstein}. There is, however, a considerable corpus of manuscripts, the so-called working sheets, which are unknown. These materials are for the most part not yet understood, or even identified, although they are open to researchers as part of the Albert Einstein Archives in Jerusalem. These sheets require careful analysis. 

The purpose of this paper is briefly to describe these documents and propose a cautious preliminary assessment of what we might learn from a more thorough future investigation. Given the fact that these working sheets are still largely unidentified, uninterpreted, and unknown, such assessment can, of course, not go into any details of Einstein's unified field theory. Instead, it is hoped that this paper may help instigate future historical work on this important aspect of Einstein's oeuvre and of theoretical physics of the twentieth centruy at large.

In the following, I will first give a brief description of the Albert Einstein Archives and of the ongoing editorial project of the \emph{Collected Papers of Albert Einstein} as the major resources for current Einstein scholarship. I will then introduce Einstein's manuscripts as a resource for historical research and, in particular, describe the still unexplored working sheets on unified field theory. I will then give a few examples of the kind of insights that could be gained from a closer analysis of these sheets.

\section{The Albert Einstein Archives, the \emph{Collected Papers of Albert Einstein}, and his Working Sheets}
\label{sec:1}

Einstein died on 18 April 1955 in Princeton, New Jersey, where he had been working at the \emph{Institute for Advanced Study} (IAS) since his emigration from Nazi Germany in October 1933. Einstein had been visiting the United States on a trip to Caltech in Pasadena, California, in early 1933 when the Nazis came to power. He returned to Europe that spring, staying mostly on the Belgian coast and visiting Switzerland and England, without however returning to Germany. He was accompanied by his second wife Elsa (1876--1936) on his return to the US in October and they were later joined by Einstein's step-daughter Margot Einstein (1899--1986). Accompanying Einstein in 1933 were also his assistant Walther Mayer (1887--1948) and his secretary Helen Dukas (1896--1982). Dukas had begun working as Einstein's secretary in 1928 in Berlin and would continue to do so for the rest of Einstein’s life. 

In fact, after his arrival in Princeton in the fall of 1933, Einstein would never leave the U.S. again, except for a brief trip to Bermuda in 1935 in order to formally begin the legal procedure of his U. S. naturalization completed in 1940 \cite[p. 452--3]{PaisA1982Subtle}. His second step-daughter Ilse Kayser-Einstein (1897--1934) and her husband Rudolf Kayser (1889--1964) arranged to have most of his personal belongings sent from Berlin to Einstein’s new home in Princeton via sealed diplomatic pouch. These private possessions included most of his books and papers \cite[p. 450]{PaisA1982Subtle}, \cite[10--13]{CalapriceAEtal2015Encyclopedia}. Therefore, the literary estate and library extant at the time of his death in 1955, both in his office at the Institute and in his home at 112 Mercer St., included books and papers from his Berlin period and even from the earlier years spent in Switzerland and Prague.

Helen Dukas and Otto Nathan became the executors of Einstein’s literary estate as per his last will. Dukas continued to work at the Institute until her own death in 1982 \cite{BuchwadD2005Project}, \cite[10--13]{CalapriceAEtal2015Encyclopedia}. For the intervening 27 years, Dukas organized Einstein's papers into what became the Albert Einstein Archives. Simultaneously, efforts to set up an editorial project to publish Einstein's \emph{Collected Papers} with Princeton University Press were successfully set in motion \cite[30--37]{CalapriceAEtal2015Encyclopedia}. The first volume as well as the outline of the entire project was being prepared by an editorial team headed by John Stachel. As part of these efforts, the entire archive then extant at the IAS was catalogued and microfilmed on the premises. Paper printouts of these microfilms were being made in Princeton and collated with the originals by Stachel and his team. This duplicate paper archive became the basis for the preparation of the first volumes of the \emph{Collected Papers}.

The Albert Einstein Archives was formally constituted after Helen Dukas's death in 1982, when Einstein's papers and books were packed up in Princeton and shipped to the Jewish National and University Library at the Hebrew University of Jerusalem, the legal heir of Einstein's literary estate and holder of the copyright to all his published and unpublished writings.

Dukas, who had been working as Einstein's secretary for 27 years, was a committed archivist, responsible for organizing his professional and personal correspondence. She also expended untold efforts to complete the collection by trying to obtain copies of outgoing letters from many individuals and institutions. She was a careful guardian of this material. But, to everyone’s surprise, when Einstein's papers were being packed up at IAS offices, a large stack of working sheets with unidentified calculations in Einstein's hand turned up, apparently having fallen somewhere behind the filing cabinets. The sheets were hastily added to the materials that were about to be shipped to Jerusalem. John Norton, then working on the premises as a postdoc in the editorial project of the \emph{Collected Papers}, recalls that Xerox copies had to be made for the editorial project before the sheets were to be packed as  no microfilm of these papers had been made. Since there was little time left and since the editors only had a day to secure a copy of these sheets for their duplicate archives, all Xerox machines available at the IAS were put to use, and everybody else who wanted to use the machines, Nobel-prize winner or not, was being turned away, a ``harrowing'' experience to the young postdoc, as Norton recalled.\footnote{John D. Norton, personal communication.} After their arrival in Jerusalem, the working sheets, too, were microfilmed and added to the Archives.

The original Albert Einstein Archives compiled by Helen Dukas and transferred to Jerusalem filled a total of 61 microfilm reels, each containing 1,000 individual page images that included explanatory cover sheets created by Dukas. The newly discovered stack of working sheets amounted to another one and a half reels of microfilm, for a total of some 1,135 pages of calculations on 808 sheets of paper for reel Number 62, and a total of some 626 pages on 416 sheets for reel Number 63. All in all, the stack thus contained a total of some 1,800 pages which were added to the Archives with archival signatures 62-001 to 63-416.\footnote{Documents in the Albert Einstein Archives are identified by archival numbers that reflect the original structure of the microfilms: a first number identifies the reel; the second number the individual image of the page in the reel. This numbering scheme has been maintained since then and is still used today for newly accessioned material, even though no microfilming is done any longer.} Either still in Princeton, or in Jerusalem, or during the transportation, the physical sequence of the pages suffered an apparently haphazard rearrangement  (see below). In Jerusalem, the working sheets were stamped with an archival number template stamp which was filled with a handwritten sequential number before the pages were microfilmed. In support of the editorial project, a hardcopy of the microfilm was printed on paper and shipped back to the editorial offices in Princeton, where they, too, were added to the duplicate archives.

It so came about, therefore, that the offices of the editorial project in Princeton acquired two huge stacks of Xerox copies and microfilm printouts, respectively, each stack containing almost 2,000 pages filled with unidentified and undated calculations by Einstein. These copies moved with the rest of the duplicate archives when the editorial offices relocated from Princeton to Boston in 1985, and from Boston to their current premises in Pasadena in 2000. Due to their origin, the stack of Xerox copies had no archival numbers, while the stack of microfilm printouts carried archival numbers added in the process of microfilming in Jerusalem.

In 2002, a preliminary analysis of these pages was undertaken at the editorial project in Pasadena. The stack of printouts carrying archival numbers was scanned and a database was created capturing basic information about each page, its physical characteristics, any peculiarities and, most significantly, a transcription of all legible, non-mathematical words. This database then was used to go through the set of Xerox copies made in Princeton; each page of the Xerox stack could then be identified with its corresponding page in the microfilm and hence with the corresponding original as they are now preserved in Jerusalem. 

The result of that preliminary examination was the following: 
\begin{enumerate}
\item[a)] Both stacks were copies of the same set of originals. Since the Xerox copies carried no archival numbers this had been far from clear, but all Xeroxed pages could be identified with extant pages in the archives.
\item[b)] The physical sequence of sheets in the two stacks differs considerably. This means that the physical sequence of the sheets as they are now preserved in Jerusalem differs from the physical sequence that the stack must have had in Princeton. But the physical sequence of the Xerox copies, which reflect the original physical sequence more closely, has been preserved and may help in the future to identify groups of related pages. 
\item[c)] We tentatively dated the sheets to a period between the late 1920s and Einstein's death in 1955. Specifically, we were able to establish that in all probability none of the material dated to a time earlier than 1928. We also established that some of the material can be dated to a narrower time-span, from the late 1920s to the fall of 1933. \end{enumerate}

After the publication in 2018 of Volume 15 of the \emph{Collected Papers} covering the period between June 1925 and May 1927 of Einstein's life \cite{CPAE15}, work began on Volumes 16 and 17, covering the years 1927 to 1931. Therefore, Einstein's working sheets have again become of interest to the editorial team. It is clear that a number of pages contain calculations that pertain to publications and correspondence from the period covered by the next two volumes of the \emph{Collected Papers of Albert Einstein}.

The sheer amount of these working sheets in addition to the lack of coherence in our current understanding of these calculations present a considerable challenge to the editors of Einstein's \emph{Collected Papers} \cite{SauerT2004Challenge}. More recently, high-quality 600dpi colored scans of all originals were produced at the Einstein Archives in Jerusalem; these images now replace the inferior quality images of the black-and-white prints \cite{MendelssonDEtal2014Project}. These scans now make it possible to decipher faint writing, analyze paper quality and other physical features, and enrich and correct the original 2002 database created for these working sheets. 

Other individual pages in the Einstein Archives that contain similar or even related calculations are also being added to the improved database of working sheets in the hopes of creating a more complete collection. Recently, some 84 sheets of calculations were added. They were part of the estate of Einstein's assistant Ernst Gabor Straus (1922—1983) and were donated to the Albert Einstein Archives through a gracious gift by the Crown-Goodman Family Foundation, thus making it possible for researchers to analyze these papers together with the rest of the Einstein papers.\footnote{Press release provided by Hebrew University of March 6, 2019; news item on phys.org of March 6, 2019.}

In their sheer numbers, Einstein's working sheets would fill three or four volumes of the \emph{Collected Papers}, not counting any editorial apparatus. On the other hand, there is reason to hope that the calculations contain interesting information that will allow us to reconstruct Einstein's ideas and give us an unobstructed, direct view of the evolution of his thinking. An analysis of these sheets will permit us to walk along with him on unsuccessful or hidden paths, dead ends, or redundant detours, but will also help us understand much of his later correspondence and experience the challenges of exploring his program of unified field theory beyond what has found its way into his published papers.

In the following I will try to indicate some of the insights that may be gained from a careful analysis of scientific working sheets, calculations and notes, both in general and with respect to the unexplored trove of documents of Einstein's search for a unified theory.

\section{What can we learn from Einstein's unpublished notes and working sheets?}

Unfortunately, no manuscripts or notes from Einstein's early years have survived. We thus have no direct documentary source for insight into his thinking leading up to his \emph{annus mirabilis} of 1905. Indeed we hardly have correspondence from those years except for the very significant exchanges with his first wife Mileva, published in the first volume of the \emph{Collected Papers} \cite{CPAE01}. 

This situation changes with respect to his search for a general theory of relativity in the decade 1905--1915. In addition to a number of publications and an extensive correspondence, we are lucky to have some unpublished notes and notebooks that allow us to reconstruct in considerable detail Einstein's path toward general relativity. Foremost among these materials is Einstein's so-called Zurich Notebook which dates from the period between summer 1912 and spring 1913 and documents Einstein's and his friend and colleague Marcel Grossmann's (1878--1936) search for a general theory of relativity.\footnote{An online facsimile version of this notebook is available at the Einstein Archives Online (www.alberteinstein.info), either in its gallery or by searching for ``Zurich Notebook''.} The significance of these notes for attempts to reconstruct how Einstein found his field equations was recognized already by John Norton \cite{NortonJ1984Einstein}. A line-by-line analysis of this notebook has indeed allowed us to get direct insight into Einstein's heuristics at this crucial period of his work \cite{RennJEtAl1999Heuristics},\cite{RennJ2007Genesis}. 

A second set of materials, dating from 1913 to 1914, contains notes that document Einstein's and his friend Michele Besso's (1873--1955) attempts to account for the anomalous advance of Mercury's perihelion on the basis of the so-called Entwurf equations, i.e.\ the gravitational field equations of the precursor theory of general relativity found by Einstein and Grossmann in 1913 \cite[Doc.~14]{CPAE04}. A careful reconstruction of these notes has given an answer to, among other things, the puzzle of how Einstein was able to compute the correct value for the perihelion advance in November 1915 within days after settling on a new set of field equations and giving up the Entwurf theory for good \cite{EarmanJEtAl1993Explanation}. 

A third notebook from those years \cite[Appendix A]{CPAE03} is less coherent but also contains notes among which some were identified to document calculations of gravitational lensing that are fully equivalent to, but predate, the corresponding calculations which Einstein published in 1936, more than twenty years later \cite{RennJEtal1997Origin}, \cite{RennJEtal2003Eclipses}. A subsequent acquisition of an unknown item of correspondence by the Einstein Archives has allowed us to identify the context of these early calculations as being related to an (unsuccessful) attempt to explain Nova Geminorum 1912 as a lensing phenomenon, a contextual explanation which also implied a probable re-dating of parts of the early notes to 1915 \cite{SauerT2008Nova}.

These examples pertain directly to Einstein's search for his general theory of relativity and thus document a heuristics of theory formation that proved uniquely successful. From studying these documents, we can learn about the dynamics of the emergence of a new theory of gravitation, of space and time, and, indeed of the universe \cite{RennJ2007Genesis}, \cite{GutfreundHEtal2015Road}. Manuscripts and calculations from Einstein's later years are less spectacular in this respect. Nevertheless, they too allow us insights into the workings of a highly creative mind. And even if for these later years we may not be studying Einstein on his royal and successful road to general relativity, we can still expect to see a clever physicist and a creative theorist at work. In what follows I will indicate a few examples of possible interest contained in Einstein's later working sheets.

Einstein kept a travel diary during his half-year long journey to Japan in the fall of 1922 \cite[Doc.~379]{CPAE13}. He left the port of Marseille on a steamer on October 7, 1922, and arrived in Japan some six weeks later on November 17, with several stops in Colombo, Singapore, Hong Kong, and Shanghai. In Japan, he spent six busy weeks traveling the country and giving lectures at various places and to various audiences. On December 29, he left Japan for his return trip. In the quiet aboard the steamer, he sat down and pondered a proposal for a generalization of general relativity, recently published by Arthur S. Eddington. After a few days, he had an idea of his own for a generalized theory along the Eddington approach and wrote a little note explaining his idea. In his diary, he wrote on January 9: ``New idea for the electromagnetic problem of the general theory of relativity.'' \cite[p.~555]{CPAE13}. On January 7 and 8, he had been ``Thinking about general relativity.'' (ibid.), and on the 9th he noted ``Writing of paper on gravitation and electricity.'' (ibid.) But before he could mail the manuscript of his paper at the next mainland stopover, he realized that he had made a mistake and needed to reconsider his new approach. For January 13 he noted in the diary: ``Discovered fly in my electricity ointment in the afternoon. A pity. True tropical heat.'' (ibid.) The following days were spent aboard ship rethinking his earlier idea and trying to find a way out of the difficulty he had discovered in his earlier manuscript. This process of revision of his theory happens to be documented by sequence of 20 pages of calculation which Einstein jotted down in the back end of his travel diary. At the conclusion of this process, he found a solution to the earlier difficulty, or so he believed, and wrote a revised version of the manuscript which he then mailed off to Max Planck (1858--1947) on his arrival in Suez, asking Planck to present it on his behalf to the Prussian Academy for publication in its Proceedings. This episode is an interesting example of Einstein's way of thinking that is also neatly and almost completely documented: We have the initial draft manuscript of 9 January 1923 \cite[DOC.~417]{CPAE13}, we have 20 pages of calculations in reaction to a difficulty that he found before sending it off \cite[Doc.~418]{CPAE13}, and we have the revised manuscript which was completed on 22 January 1923 and which was eventually published \cite{EinsteinA1923Relativitaetstheorie}, \cite[Doc.~425]{CPAE13}. Together with further hints from his diary and correspondence, we have here a prime example of documentation of a thought process, beginning with a new idea and the working out of a theory, the realization of a difficulty, the reaction to the difficulty, its eventual resolution by a revision of the original idea within the same heuristic setup, and the final condensation of the idea into a published paper.\footnote{T.~Sauer, publication in preparation.}

This episode is indeed an early example of a pattern that would repeat itself many times in Einstein's subsequent intellectual career and search for a unified field theory. In this particular instance, we are fortunate to have as complete a documentation of this episode as can be hoped for and, since the calculations are found in the back of a \emph{bound} notebook, we are in the fortunate position that the \emph{logical} and \emph{chronological} sequence of the calculations can be assumed to coincide with their \emph{physical} sequence. Furthermore, we can date those calculations most precisely and can reasonably assume that they represent a complete and coherent train of ideas.

With the large stack of Einstein's working sheets from his later years we are fortunate to have extensive documentation of his thinking and theorizing during these years. Unfortunately, these sheets are less coherent and not neatly bound together like the pages in the back of his Japan travel diary which otherwise very much resemble the later working sheets. In fact, we know that the working sheets of reels 62 and 63 of the Albert Einstein Archives are out of sequence and that in many instances they may also be incomplete. A reconstruction of the flow of ideas crystallized in these pages therefore poses a considerable challenge to historians and Einstein scholars. Nevertheless, we can hope to get surprising insights into the working of Einstein's mind. Let me mention two examples.

Most of the pages apparently document Einstein’s various approaches in his search for a unified field theory. But there are also some that deal with problems in classical general relativity, and there are a few that deal with problems in other areas of physics. As an example for the latter category, the stack contains a single page with a brief formulation of the EPR paradox by Einstein (AEA 62-575r). It is one of the few formulations in Einstein's own hand of the famous argument published in one of his most frequently cited papers \cite{EinsteinAEtAl1935Description}. Such authenticity is all the more valuable since we know that Einstein did not himself compose the version of the published EPR paper and did not fully approve of it. Instead he complained about its unnecessary eruditeness which would bury the main idea. It is also the only version of the argument by Einstein which formulates the EPR paradox in terms of spin variables, as was first done in print by David Bohm (1917--1922), and a dating of the manuscript indicates indeed temporal proximity to that time and suggests that Einstein's version was phrased in reaction to his interactions with Bohm. Nevertheless, the way the argument is phrased poses a quandary, since Einstein seems to think about the paradox in a way that is different from and, in fact, not readily compatible with, our current understanding of the EPR argument. For a detailed discussion of this particular page, see \cite{SauerT2007Manuscript}.

As an example of problems from classical general relativity, the stack also contains four pages of calculations that deal with gravitational lensing and that are closely related to Einstein's early calculations mentioned above and to his 1936 publication. A careful analysis and reconstruction of these additional manuscript pages on gravitational lensing has allowed us to identify what we call a ``space of implications'' for different routes of exploring the concept of gravitational lensing \cite{SauerSchuetz2019Exploring}. The observation here is that the same conceptual idea can result in very similar actual calculations. These are related by its overall aim and generic technical means but differ in details like notation, or the sequence in which calculational steps like computing an explicit solution, or taking various approximations, are being performed. The final result is predetermined by the heuristic setup, but we see Einstein arrive at equivalent expressions for the lensed image following different routes in this space of possible pathways.

\section{Outlook}

The editorial project of the \emph{Collected Papers} is currently preparing the next two volumes of Einstein's writings and correspondence for the period June 1927 to May 1931. Einstein’s thinking was dominated during these years by an approach to unified field theory that he called distant parallelism, which he explored between summer 1928 and summer 1931 \cite{DebeverR1979Letters}, \cite[pp. 234--258]{VizginV1994Theories}, \cite{GoldsteinCEtal2003Varieties}, \cite{GoennerH2004History}, \cite{SauerT2006Equations}, \cite{SauerT2014Program}. There are some 100 pages of Einstein's working sheets that clearly relate to this episode. On the other hand, the episode again resulted in almost a dozen papers during that period, and Einstein's correspondence, in particular, extensive exchanges of letters with Herman Müntz (1884--1956), Roland Weitzenböck (1885--1955), Jakob Grommer (1879--1933), Cornelius Lanczos (1893--1974), Elie Cartan (1869--1951), Walther Mayer, and others document Einstein at work. In this correspondence we repeatedly find suggestions as to what should be done and investigated, or hints at ideas that should be followed up, or calculations that should be done---only to find unanticipated difficulties. A better understanding of this episode requires a close analysis of both the published papers and Einstein's correspondence from that time, in close connection with an attempt at reconstructing his calculations. This is a considerable challenge, but it also offers the perspective of being able to understand Einstein's thinking and, in fact, physical theorizing in the middle of the twentieth century, much closer to the actual documents than what  publications alone would ever allow us to do. Such direct access to Einstein's world of ideas may also help today's physicists who explore approaches like that of teleparallel gravity again from a modern point of view, without having to rely on received opinions about Einstein's alleged stubbornness or futile attempts along the program of a unified theory.

\bigskip
\emph{Acknowledgment}. I wish to thank Diana Kormos Buchwald for encouragement and helpful comments on an earlier version of this paper.

%
\nocite{*}
\bibliographystyle{apalike}
\bibliography{epjhuft}


\end{document}